\documentstyle[prl,aps,tighten,epsf]{revtex}

\def\la{\mathrel{\mathpalette\fun <}}
\def\ga{\mathrel{\mathpalette\fun >}}
\def\fun#1#2{\lower3.6pt\vbox{\baselineskip0pt\lineskip.9pt
\ialign{$\mathsurround=0pt#1\hfill##\hfil$\crcr#2\crcr\sim\crcr}}}

\def\rn{}
\def\nn#1 #2{#2. #1}                            
\def\nnn#1 #2 #3{#2. #3. #1}                    
\def\nnnn#1 #2 #3 #4{#2. #3. #4. #1}            
\def\nnnnn#1 #2 #3 #4 #5{#2. #3. #4 #5. #1}     

\def\rf#1;#2;#3;#4;#5 {{\frenchspacing\par\rn#1, #3 {\bf #4}, #5 (#2). \par}}
\def\rfbook#1;#2;#3;#4;#5 {{\frenchspacing\par\rn#1, {\it #3} (#5, #4, #2).\par}}
\def\rfprep#1;#2;#3 {{\par\frenchspacing\rn#1, #3 (#2).\par}}

\begin{document}

\twocolumn[
\hsize\textwidth\columnwidth\hsize\csname@twocolumnfalse\endcsname

\title{Bounds on the possible evolution of the Gravitational Constant
      \\ from Cosmological Type-Ia Supernovae}

\author{E. Gazta\~naga$^{\rm a,b}$,  
        E. Garc\'\i a-Berro$^{\rm c,b}$,  
        J. Isern$^{\rm d,b}$,
        E. Bravo$^{\rm e,b}$ \& 
        I. Dom\'\i nguez$^{\rm f}$}

\address{$^{\rm a}$ INAOE, Astrof\'{\i}sica, Tonantzintla, Apdo.  Postal
        216 y 51, 7200, Puebla, Mexico}
\address{$^{\rm  b}$ Institut d'Estudis  Espacials de Catalunya, Edifici
        Nexus, Gran Capit\'an 2-4, 08034 Barcelona, Spain}
\address{$^{\rm  c}$  Departament de F\'{\i}sica  Aplicada,  Universitat
        Polit\`ecnica  de Catalunya,  Jordi Girona  Salgado s/n, M\`odul
        B--5, Campus Nord, 08034 Barcelona, Spain}
\address{$^{\rm  d}$  Institut de Ciencies  de l'Espai  (CSIC)}
\address{$^{\rm  e}$  Departament de  F\'{\i}sica i Enginyeria  Nuclear,
        Universitat  Polit\`ecnica  de  Catalunya,  Avda.  Diagonal 647,
        08028 Barcelona, Spain}
\address{$^{\rm f}$  Departamento de  F\'\i sica Te\'orica y del Cosmos,
        Universidad de Granada, 18071 Granada, Spain}

\bigskip
~~

\date{April 2001}
~
\maketitle

\begin{abstract}  
  
Recent  high-redshift Type Ia supernovae  results can be used to set new
bounds on a possible  variation of the  gravitational  constant $G$.  If
the local value of $G$ at the space-time  location of distant supernovae
is different,  it would change both the kinetic  energy  release and the
amount of $^{56}$Ni synthesized in the supernova outburst.  Both effects
are related to a change in the  Chandrasekhar  mass  $M_{\rm  Ch}\propto
G^{-3/2}$.  In addition, the integrated variation of $G$ with time would
also affect the cosmic  evolution and therefore the luminosity  distance
relation.  We show that the later  effect in the  magnitudes  of Type Ia
supernovae is typically  several times smaller than the change  produced
by  the   corresponding   variation  of  the   Chandrasekhar   mass.  We
investigate  in a  consistent  way how a varying  $G$ could  modify  the
Hubble  diagram of Type Ia supernovae  and how these results can be used
to set upper bounds to a hypothetical  variation of $G$.  We find $G/G_0
\la 1.1$ and $\dot G/G \la  10^{-11}  {\rm  yr}^{-1}$  at  redshifts  $z
\simeq 0.5$.  These new bounds extend the currently available constrains
on the evolution of $G$ all the way from solar and stellar  distances to
typical scales of Gpc/Gyr, i.e.  by more than 15 orders of magnitudes in
time and distance.

\end{abstract}


\draft

\bigskip

\pacs{PACS numbers: 98.80.Cq, 04.50.+h}

]

\section{Introduction}

One  of  the  most  important   challenges  of  modern  physics  is  the
quantization  of the  gravitational  force.  The undergoing  attempts to
create such theories has  re-opened  the subject of varying  fundamental
constants.  To this regard it is worth  noticing  that the  constancy of
the  fundamental  constants,  and  of  the  gravitational   constant  in
particular,     has    been     questioned     for    a    long     time
\cite{milne1,milne2,dirac,jordan1,jordan2}  and that early  attempts  to
unify gravity with electromagnetism  \cite{kaluza,klein}  predicted such
kind of variations.  Although  modern  theories,  like the string theory
and  the  M-theory  (see  \cite{daniel}  for a  recent  review),  do not
necessarily  require  a  variation  of the  fundamental  constants  they
provide a natural and self-consistent framework for such variations (see
\cite{bbH2O}  and  \cite{murphy}  for  excellent   descriptions  of  the
theoretical  background).  As a general result, modern theories  predict
that in the ordinary  three-dimensional  subspace, gauge  couplings like
the fine structure  constant $\alpha$ or the gravitational  constant $G$
should  vary as the  inverse  square  of the  mean  scale  of the  extra
dimensions.  Hence, the  evolution  of the scale size of the  additional
dimensions  is  related  to  the  variation  of  fundamental   constants
\cite{marciano,Barrow87,DaP}.  Moreover  it  has  been  recently   shown
\cite{DaP}  that a  cosmological  variation  of $\alpha$  may proceed at
different rates at different  locations in space-time.  The way in which
the time  variations  of $\alpha$ and $G$ are linked is model  dependent
but a  typical  relation  is:  $\Delta\alpha/\alpha^2  \sim\Delta  G/G$.
There have been  several  attempts to measure the rate of  variation  of
$\alpha$,  providing  different  results for different  look-back times.
For instance,  \cite{avelino}  used the recently released CMB anisotropy
data to set up early-universe constraints on a time-varying $\alpha$ and
found no evidence  for such a change,  whereas  \cite{murphy,webb}  used
high  resolution   spectroscopy  of  QSO  absorption   systems  to  find
statistical  evidences  for  a  smaller  $\alpha$  at a  redshift  range
$0.5<z<3.5$.  Of course,  since a  cosmological  variation  of  $\alpha$
(and, consequently, of $G$) can proceed at different rates for different
redshifts  \cite{DaP}  both  studies are not  necessarily  in  conflict.
There have been also many  attempts to measure a time  variation  of the
gravitational  constant which will be discussed  later in \S V.  For the
moment it is important  to mention  here that most of these  bounds come
either from local  measurements  (the sun, our solar system or the solar
neighborhood)  or from very early times  measurements  (namely  Big-Bang
nucleosynthesis),  whereas at intermediate look-back times there are not
such measurements.

Type Ia supernovae (SNIa) are supposed to be one of the best examples of
standard  candles.  This  is  because,  although  the  nature  of  their
progenitors  and the  detailed  mechanism  of  explosion  are  still the
subject  of a  strong  debate,  their  observational  light  curves  are
relatively well understood and their  individual  intrinsic  differences
can be accounted for.  Under these assumptions, thermonuclear supernovae
are well suited  objects to study the Universe at large,  especially  at
high redshifts $(z\sim 1.0)$, where the rest of standard candles fail in
deriving   reliable   distances,  thus  providing  an  unique  tool  for
determining  cosmological  parameters or discriminating  among different
alternative cosmological theories.

Using the observations of high redshift Type Ia supernovae $(z>0.1)$ and
low redshift $(z<0.1)$  supernovae, both the Supernova Cosmology Project
\cite{per} and the High-$z$  Supernova Search Team \cite{rei} found that
the peak  luminosities  of distant  supernovae  appear to be $\sim 0.20$
magnitude  fainter  than  predicted  for  an  empty  universe  and  0.25
magnitude  fainter than predicted by a standard  decelerating  universe,
with a presumed mass density  $\Omega_M\simeq 0.3$.  To be more precise,
at the  $1\sigma$  confidence  level, the results of both  groups can be
well   approximated   by  the   relation:
  
\begin{equation}
0.8  \,  \Omega_M   -0.6  \,
\Omega_\Lambda  = -0.2  \pm  0.1.
\label{omegasnIa}
\end{equation}

However  these  conclusions  rely on the  assumption  that  there  is no
mechanism able to produce an evolution of the observed light curves over
cosmological  distances.  In other words:  both teams  assumed  that the
relation  between the intrinsic peak  luminosity  and the time scales of
the light  curve  were  exactly  the same for both the  low-$z$  and the
high-$z$ supernovae.  The possible consequences for evolutionary effects
in SNIa due to  changes  in the zero age  mass  and  metalicity  of the
progenitor    star   have   been    explored    by    several    authors
\cite{Inma99,Hoflich20,Inma01}, who found that changes in the underlying
population  cause a change in the maximum  brightness  by about  0.1-0.2
magnitudes.

The SNIa results have already  motivated a significant  number of papers
that  search  for  bounds  on the  variation  in  fundamental  constants
\cite{bm00,ame,GB99,pr00,barrow01}  or new cosmological  scenarios, such
as  quintessence   models   \cite{wang}  and  scalar-field   cosmologies
\cite{waga00}.  This burst of interest is due to the conceptual problems
that arise  from  infering  the  existence  of a  cosmological  constant
$\Lambda  \simeq  10^{-122}  c^3/G/\hbar$  or facing the  cosmic  (dark)
matter problem (see \cite{tegmark} and references  therein).  There have
been many suggestions that the apparent  complications that arise can be
eliminated      by      modifying      the      laws     of      gravity
\cite{Milgrom83,Milgrom98,Damour99,Mann00,Saini00,Bois00,ep01,Gazta00}.

Recent cosmological  observations, such as the lastest CMB Boomerang and
Maxima data \cite{boo,max} indicate a flat Universe:  $\Omega_R=0$, i.e.
$\Omega_M+\Omega_\Lambda=1$.  This  result,   together  with  the  above
equation   \ref{omegasnIa}   points  in  the  direction  of  a  non-cero
$\Lambda$,   although   other    interpretations   are   also   possible
\cite{McGaugh00,bgss01}.

The purpose of this paper is to analyze the effect of varying $G$ in the
current  interpretation of the Hubble diagram of distant SNIa and to use
this  analysis to set upper  bounds on its rate of change.  The paper is
organized as follows:  in \S II we describe the effects of a varying $G$
on the physics of supernovae;  in \S III and in the Appendix, we analyze
the  effects of a varying  $G$ on the  luminosity  distance  of  distant
supernovae.  In \S IV we present a likelihood  analysis of the SNIa data
which is then used to set upper bounds in the evolution of $G$.  Finally
in section V we discuss our results and draw our conclusions.

\section{The effects of a varying $G$ on the physics of supernovae}

Simple analytical models of light curve (see, for instance,  \cite{arn})
predict that the peak luminosity is  proportional  to the mass of nickel
synthesized, which in turn, to a good approximation, is a fixed fraction
of the  Chandrasekhar  mass  $(M_{\rm  Ni}\propto  M_{\rm  Ch})$,  which
depends on the value of gravitational  constant as:  $M_{\rm  Ch}\propto
G^{-3/2}$.  The actual  fraction  varies when  different  specific  SNIa
scenarios  are  considered  (e.g.  \cite{kmh93,ggij}),  but the physical
mechanisms relevant for type Ia supernovae  naturally relates the energy
yield  to  the   Chandrasekhar   mass.  Here  we  will  only   focus  on
Chandrasekhar  mass models  since there are growing  evidences  that the
sub-Chandrasekhar mass models do not fit well the observations (see, for
instance,  \cite{Branch}).  In summary, whatever the actual scenario is,
we will  assume  that  the  same  mechanism  for  the  ignition  and the
propagation  of the  burning  front is valid  for  SNIa at high  and low
redshifts.  Thus, since the peak luminosity is proportional to the total
amount of nickel synthesized in the supernova outburst we have $L\propto
G^{-3/2}$, and, therefore, for a slow decrease of $G$ with time, distant
supernovae  should be dimmer  than  predicted  for a standard  scenario.
Under this assumptions, we have:

\begin{equation}
M-M_0=\frac{15}{4}\log\Big(\frac{G}{G_0}\Big)
\label{peak}
\end{equation}

\noindent  where $M$ stands, as usual, for the absolute  magnitude,  $G$
for the precise value of the gravitational  constant at a given redshift
and the  subscript 0 denotes  their  local  values.  Note that the above
equation  does not require  knowledge of the  (unknown)  proportionality
constant relating the supernova  luminosity with the Chandrasekhar mass.
This  dependence  factorizes  out under the  assumptions  above, and the
final differential  result is only sensitive to the values of $G$.  From
this equation we can see that in order to reduce the apparent luminosity
of distant  supernovae by $\Delta m \simeq 0.2$ a dramatic change of $G$
is required:  $G/G_0\simeq  1.13$.  This value  should be regarded as an
upper bound in the sense that part or all of the  $\Delta m \simeq  0.2$
difference  found  by  \cite{per,rei}   could  be  attributable  to  the
hypothesis of an accelerating universe.

In order to test the validity of our argument we have  computed a series
of models of Type Ia supernovae  explosion and their corresponding light
curves,  according to the  procedure  described  in  \cite{bravo99}  and
references  therein,  with the present  local  value of $G$, 1.1 and 1.2
times this value.  The explosion model was a delayed detonation starting
from a central  density of  $2.0\times  10{^9}\,  {\rm  g/cm}^3$  a core
temperature  of  $2.0\times  10^8$~K  and  making  the  transition  from
deflagration to detonation when the flame density went below  $2.0\times
10^7\, {\rm g/cm}^3$.  Our study is based on delayed detonation  models,
because  these have been found to  reproduce  the optical  and  infrared
light  curves  and  spectra  of  Type  Ia  supernovae   reasonably  well
\cite{Hoflich20,Hof95,Hof95b,Hof96,Nug97,Lentz00,Ger20}.    The    model
parameters,  ignition  density and  transition  density,  are those that
allow us to  reproduce a typical  Type Ia  Supernovae.  The  results are
shown in Table 1 and Figure 1.

\begin{figure*}
\epsfxsize 3.4in 
\epsfbox{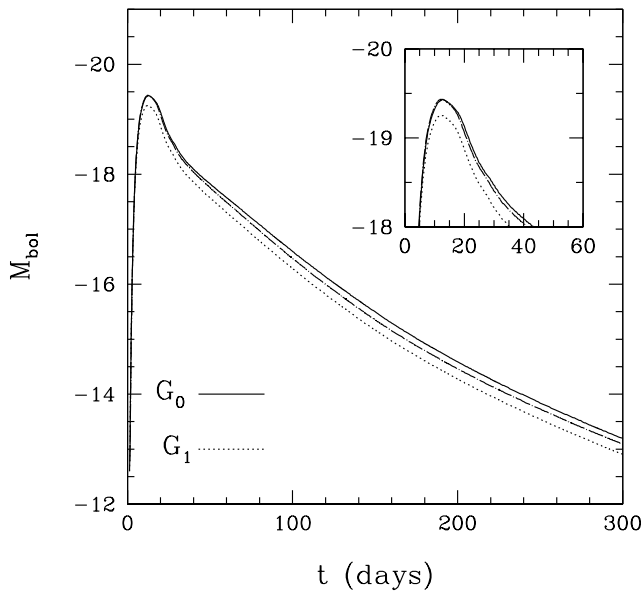}
\caption[]{Bolometric  light  curves of SNIa for the local  value of the
gravitational  constant,  $G_0$ (solid line), for  $G_1=1.1G_0$  (dotted
line) and for $G_1$  shifted  upwards by 0.18  magnitudes.  See text for
details.}
\label{fig1}
\end{figure*}

In Table 1 we show the mass of the  white  dwarf  model  in  hydrostatic
equilibrium  from which the explosion was  computed,  $M_{\rm  WD}$, the
kinetic energy, $K$, the mass of synthesized  nickel,  $M_{\rm Ni}$, the
peak bolometric  magnitude,  $M_{\rm bol}^{\rm  peak}$, and a measure of
the width of the  light  curve,  $\Delta  m_{15}$  --- see  below  for a
precise  definition.  All the masses are  expressed in solar  units.  In
Figure 1 we show the light  curves for  $G/G_0=1.0$  --- solid  line ---
$G_1=G/G_0=1.1$  --- dotted line --- and this last light  curve  shifted
upwards by a constant amount of 0.18 magnitudes ---  dashed-dotted  line
--- in accordance with the behavior predicted by Eq.[\ref{peak}].

As it can be seen in Table 1 the  simple  energetic  argument  presented
above is in good agreement with the detailed calculations presented here
and, thus, the energy liberated in the supernova  outburst indeed scales
as $G^{-3/2}$.  Hence, the bolometric magnitude at the peak of the light
curve  computed with a larger value of $G$ should be moved  upwards by a
fixed constant amount which depends on the exact value of $G$.  Figure 1
clearly shows this  behavior.  It should be stressed at this point that
our  analysis  does  not  depend  on the  detailed  physics  of  Type Ia
supernovae  since the functional  dependence on the  Chandrasekhar  mass
comes from basic  physical  arguments and it is  unavoidable,  no matter
which are the (unknown)  details of the  explosion  unless we change the
physics underlaying the Chandrasekhar mechanism.

Figure 1 also shows that once this vertical  shift is done, the duration
of the supernova  outburst is also  modified by a varying $G$, being the
decline faster for the models with larger $G$, especially at late times.
As it can be seen there, in the region near the maximum  (see the insert
for a close up of this region) the difference  between both light curves
is small  when  due  account  of the  vertical  shift  is  done.  A good
parameterization  of the slope  (and,  thus, of the  width) of the light
curve is $\Delta  m_{15}$,  which is defined  as the  difference  in the
apparent  magnitude  15 days  after the  maximum.  It turns out that the
maximum of the light curve for the models presented in Fig.  1 occurs at
13.0 and 12.5 days, being $\Delta  m_{15}=0.85$ and 0.89,  respectively.
This, in turn,  implies  that since the  template  light  curve  used to
calibrate  the  distances to distant  supernovae  takes into account the
width of the light curve, and in particular  $\Delta m_ {15}$ is used, a
variation of the gravitational constant should ultimately affect as well
the final  value of the  derived  distances.  However,  given the slight
variation  of the  $\Delta  m_  {15}$  parameter  this  can  be  clearly
considered as a second order effect.  Nevertheless, as it can be clearly
seen in Fig.  1,  although  there  is not any  large  difference  in the
rise-times  or in the early time light curve due to a varying  $G$ there
is indeed an  appreciable  difference  in the  overall  duration  of the
supernova event.  To be precise the widths at $M_{\rm  bol}=-16$ are 126
and 112 days, respectively.

This  result  can be  interpreted  in  terms  of  very  simple  physical
considerations,  and in particular, in terms of the simplified  model of
light curve presented in \cite{arn} which has a reasonable  accuracy (of
the order of $\sim  20$\%).  According to this  analytic  model of light
curve, the width of the peak of the light curve of SNIa is given by:

\begin{table}
\caption{Overall  characteristics of the supernova explosion for several
values of $G$, and the same ignition density.}
\begin{tabular}{lccccc}
$G/G_0$  & $M_{\rm WD}$ 
         & $K/10^{51}\,{\rm erg}$ 
         & $M_{\rm Ni}$ 
         & $M_{\rm  bol}^{\rm  peak}$
         & $\Delta  m_{15}$\cr
\hline
1.0  &1.37  &1.34   &0.69     &$-19.43$ &  0.85 \cr
1.1  &1.19  &1.14   &0.57     &$-19.25$ &  0.89 \cr
1.2  &1.04  &0.95   &0.49     &$-19.13$ &  0.91 \cr
\end{tabular}
\end{table}

\begin{equation}
\tau\propto\Big(\frac{M_{\rm ej}^3}{K}\Big)^{1/4}
\label{time}
\end{equation}

\noindent  where $M_{\rm ej}$ is the ejected mass and $K$ is the kinetic
energy.  Within our current  knowledge of the mechanisms of explosion of
SNIa both the  ejected  mass and the  supernova  kinetic  energy  can be
considered proportional to the Chandrasekhar mass, and therefore we have
$\tau\propto M_{\rm Ch}^{1/2}$ or, equivalently, $\tau\propto G^{-3/4}$.
Thus we have

\begin{equation}
\Big\langle\frac{\tau}{\tau_0}\Big\rangle\simeq
\Big\langle\frac{G}{G_0}\Big\rangle^{-3/4}.
\label{width}
\end{equation}

The  ratio  of the  durations  of the  supernova  outburst  calculations
presented  above  matches  reasonably  well the  behavior  predicted  by
Eq.[\ref{width}].  Hence, in the case in which a  varying  gravitational
constant is considered, the overall time scales of the supernovae  light
curves should depend as well on the actual value of $G$.

It has been  recently  claimed  that  there is a mean  evolution  in the
rise-times   of  local  and   distant   supernovae   \cite{r9a,r9b}.  In
particular  the  widths of the light  curve  when the  supernova  is 2.5
magnitudes  fainter than the peak  luminosity  was found to be $\tau_0 =
45.0 \pm 0.15$ (at  $z\simeq  0$) and $\tau = 43.8 \pm 0.40$ days (at $z
\simeq 0.5$), were the errors in the widths were ascribed  solely to the
errors  in  the  rise-times.  Using  this  data  and   Eq.[\ref{width}],
\cite{GB99}  obtained  $G/G_0\le  1.037\pm 0.017$ ($2\sigma$  errors) at
$z\simeq  0.5$, a  variation  of $G$ very  similar to the one  needed to
explain   the  change  in  the  peak   luminosity.  Subsequent   studies
\cite{ald00} have demonstrated  that the rise-times of local and distant
supernovae   are   consistent   each  another  and  that  the  rise-time
uncertainties  were  underestimated,  being revised upwards to $\pm 1.2$
days  statistical and  $^{+3.6}_{-1.9}$  days due to  systematical  bias
under  extreme  situations.  According  to this last  analysis  there is
still  some room for  deviations  of the  light  curve of high  redshift
supernovae  at {\sl late  times}.  This is exactly  what we have  found.
However, it is worth  mentioning  at this point that  according  to this
study these late time deviations  systematically influence the rise-time
determinations.  Conversely,  should we have  tried to fit the late time
light  curve  we  would  have  found  a  significant  difference  in the
corresponding  rise-times.  In the light of this new analysis,  there is
no significant evidence for a possible change in $G$.  For our purposes,
since the  situation  is not yet clear from the  observational  point of
view  our   analysis   will  only  rely  on  the   limits   imposed   by
Eq.[\ref{peak}].  Note  however  that, on the other hand,  should a firm
estimate of the maximum  value of the  difference  between the local and
distant  supernovae  timescales  could  be  eventually  obtained  a very
stringent  upper  limit to the rate of  variation  of the  gravitational
constant would be derived, which,  additionally,  {\sl would not depend}
on the adopted  cosmological  model but on the well calibrated  relation
between the duration of the supernova outburst and its peak magnitude.

\section{The Hubble diagram}

The  next  question  we want  to  address  here  is how an  hypothetical
variation of the gravitational constant (of the amount of a few percent)
translates in the Hubble diagram of distant  supernovae.  In addition to
the change in the  intrinsic  energy  release and of the duration of the
supernova event induced by the variation in the Chandrasekhar  mass, the
integrated  evolution  of $G$ with time  would  also  affect  the cosmic
evolution and,  therefore, the  luminosity  distance  relation.  To make
this quantitative we need to consider non-standard  cosmological models.
It is  however  beyond  the  scope of this  paper  to  discuss  specific
theoretical models to replace the standard theory of General Relativity.
This  has  been  discussed  in  detail   elsewhere  in  the   literature
\cite{dan,de96a,de96b,wil,wil94,wil2},  and  also in the  context  of an
accelerating universe \cite{ep01,de98,beps99,ss01}.

\begin{figure*}
\epsfxsize 3.4in 
\epsfbox{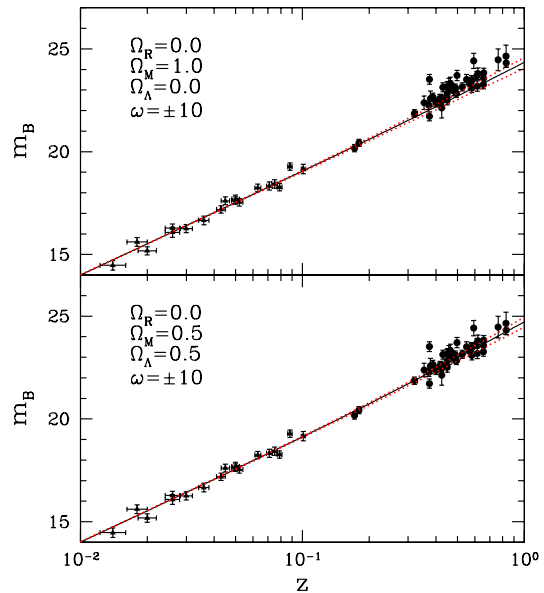}
\caption[]{Hubble   diagrams  for  the  high-redshift   SNIa  data  with
different  choices of  cosmological  parameters and different  values of
$\dot G$ .  Continuous lines  correspond to the standard  cosmology with
$G(z)=G_0$.  Dashed  lines  bracket the  effects on the peak  luminosity
from a $\simeq 5\%$ variation in $G$.  The  observational  data has been
obtained from \cite{per}.}
\label{fig2}
\end{figure*}

In the  appendix  we show in a  self-consistent  way that for  plausible
models which  incorporate a varying $G$, such as scalar-tensor  theories
(STTs),  the  possible  effect  of a  varying  $G$ on  the  cosmological
evolution gives a contribution  to the luminosity  distance  relation at
the distances of interest which is several times smaller than the effect
produced by the same  variation  of $G$ on the  Chandrasekhar  mass.  We
will therefore concentrate our analysis on the effects of the physics of
the     supernovae.     This     complements     the     analysis     of
\cite{ep01,beps99,ss01}  which  neglected  the effects in the physics of
the supernovae and used the evidence for the accelerating  universe as a
way to constraint cosmic evolution in non-standard  theories of gravity.

In analogy to STTs, we will  parameterize  the evolution of $G$ in terms
of the strength of coupling parameter, $\omega$, as:

\begin{equation}
G(z) \equiv G_0 (1+z)^{1\over{1+\omega(z)}},
\label{G(w)}
\end{equation}

\noindent  which provides our  definition  for  $\omega(z)$  --- see the
appendix  for  further  details.  Thus a value  of  $\omega  \simeq  10$
produces a $\simeq 4\%$ increase in $G$ at $z \simeq 0.5$, while $\omega
\simeq -10$  produces a $\simeq 5\%$  decrease in $G$ at $z \simeq 0.5$.
In  summary,  Eq.[\ref{G(w)}]  gives the change in $G$ as a function  of
$\omega$  while  Eq.[\ref{H(z)}-\ref{omegahat}]  give the  corresponding
cosmic evolution.  Thus we have:

\begin{equation}
m(z)=M_0 +5~\log d_L +25
+ \frac{15}{4(1+\omega)}\log\big(1+z\big) 
\label{m(z)}
\end{equation}

\noindent  where $d_L= d_L(z, \Omega  _M,\Omega  _\Lambda)$  is obtained
from  the  line-of-sight  comoving  distance  (see  the  appendix).  For
illustrative purposes Figure \ref{fig2} shows the above relation for two
representative cosmological models, including the effects of $\omega$ in
$d_L$, for $\omega= \pm 10$ (dotted lines), which correspond to a change
of $G$ of $\sim 5\%$, and the  standard  $(\omega=\infty)$  case  (solid
line).

\section{Bounds on $\dot G/G$}

We have  re-done the  likelihood  analysis  of the  Supernova  Cosmology
Project allowing for a varying $G$.  Thus, we use the same observational
data but with the magnitude-distance  relation given by Eq.[\ref{m(z)}].
Here we  have  one  extra  function  $\omega(z)$  to be  fitted.  At low
redshifts  the  last  term  in  the  r.h.s.  of  Eq.[\ref{m(z)}]  has  a
negligible  contribution.  Given that most of the SNIa at high  redshift
cluster  around $z \simeq 0.5$ and that we are expecting  $\omega$ to be
large  (so  that $G  \simeq  G_0$ in  Eq.[\ref{G(w)}]),  the  effect  of
$\omega(z)$ in the fit to  Eq.[\ref{m(z)}]  is dominated by the value of
$\omega$  at the mean  redshift  of the SNIa  sample,  $z  \simeq  0.5$.
Hence, we can approximate $\omega(z) \simeq \omega(z\simeq 0.5)$ and fit
the data as a  function  of this new  extra  parameter,  $\omega(z\simeq
0.5)$ or $G(z\simeq 0.5)/G(z \simeq 0) \equiv G/G_0$.

Our results are plotted in Figures 3 and 4, where we show the confidence
contours  (at the  99\%,  90\%,  68\% ---  solid  lines  --- 5\% and 1\%
confidence  level --- dotted lines)  obtained from the fit to the Hubble
diagram  of SNIa.  Figure 3 shows the  likelihood  contours  for  $G/G_0
\equiv G(z\simeq  0.5)/G(z \simeq 0)$ as a function of  $\Omega_\Lambda$
for a flat $\Omega_R=0$  universe, whereas Figure 4 shows the likelihood
contours in the  $(G/G_0,\Omega_M)$  plane for the case $\Omega_ \Lambda
=0$.  As it can seen in these figures, the expected  departures from the
standard case $(G/G_0 \simeq 1)$ are quite small for a reasonable choice
of cosmological  parameters.  This, in turn, justifies our approximation
$\omega(z) \simeq \omega(z\simeq  0.5)$.  It is also worth mentioning at
this point that we have also  tried  linear  fits to  $\omega(z)  \simeq
\omega_0 + \omega' z$ and found equivalent results.

\begin{figure}
{\epsfxsize 3.4in
\epsfbox{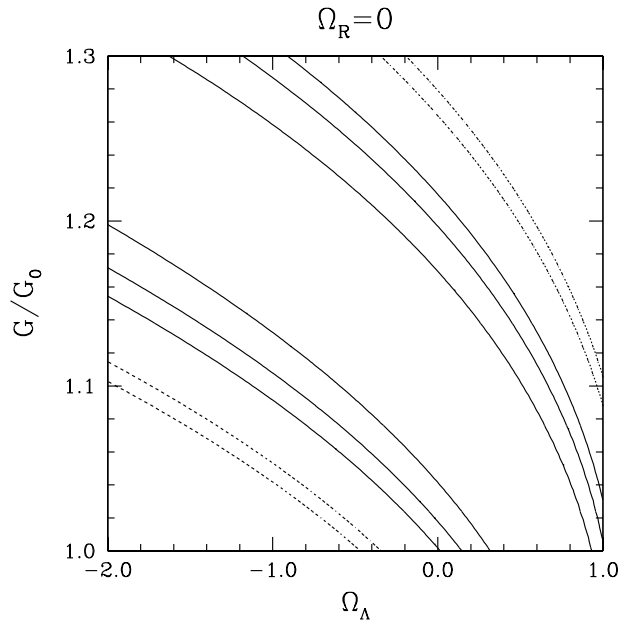}}
\caption[junk]{Confidence  contours  for the best fit SNIa  data in the
plane $(G/G_0, \Omega _{\Lambda})$ for a flat universe $\Omega_R=0$.}
\label{fig3}
\end{figure}

\begin{figure}
{\epsfxsize 3.4in
\epsfbox{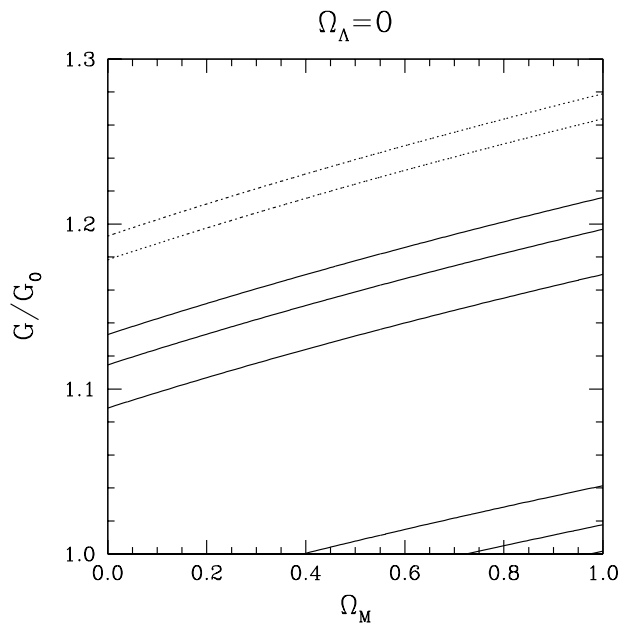}}
\caption[junk]{Confidence contours for the best fit SNIa data
in the plane $(G/G_0, \Omega _{M})$
for $\Omega_\Lambda=0$.}
\label{fig4}
\end{figure}

The  confidence  intervals of these figures can be used to set bounds in
$G/G_0$.  These are  bounds  in the  sense  that,  given a  cosmological
model,  we  assume  that  all the  difference  in  SNIa  corrected  peak
luminosities can be attributed  solely to a difference in  Chandrasekhar
mass.  For example, for the flat $\Omega_\Lambda \simeq 0$ model we have
at 1$\sigma$ confidence level:

\begin{equation}
{G \over{G_0}} \la 1.2 ~~ ; ~~ \Omega_\Lambda
\simeq 0.0  ~,~ \Omega_M
\simeq 1.0,
\label{boundl0}
\end{equation}

\noindent while for the flat $\Omega_\Lambda \simeq 0.8$ case

\begin{equation}
{G \over{G_0}} \la 1.08 ~~ ; ~~ \Omega_\Lambda
\simeq 0.8  ~,~ \Omega_M \simeq 0.2
\label{boundk0}
\end{equation}

\noindent also at $1\sigma$  confidence  level.  In terms of the $\omega
\simeq  \omega(z\simeq 0.5)$ parameter these later bounds translate into
$\omega \ga 1.2$ for $\Omega_\Lambda  \simeq 0$ and $\omega \ga 4.3$ for
$\Omega_\Lambda \simeq 0.8$.  It is important to mention here that these
constraints  are quite loose when  compared  with the bounds on $\omega$
from very long baseline  interferometry in the local Universe, $\omega_0
> 2600$,  \cite{wil2}, but, on the other hand, are new in the sense that
they correspond to an earlier (or equivalently  more distant)  Universe.
Moreover, the SNIa data provide more  interesting  limits to $\dot G/G$.
To obtain them we can use

\begin{equation}
{\dot G \over G} \simeq \left(1-{G_0\over{G}}\right) (\Delta t)^{-1}
\end{equation}
where $\Delta t$ is the look-back time to $z \simeq 0.5$:
\begin{equation}
\Delta t = \int_0^z \frac{dz'}{(1+z')\,H(z')}
\label{lookback}
\end{equation}

\noindent   with   $H(z)$   given   by   Eq.[\ref{H(z)}].  For  a   flat
$\Omega_\Lambda  \simeq  0$ model we have  that $ \Delta  t  \simeq  4.6
\times 10^9 {\rm  yr}/h_{70}$  ($h_{70}$ is the Hubble constant in units
of $H_0= 70~ {\rm Km/s/Mpc}$), while for the flat $\Omega_\Lambda \simeq
0.8$ case we have $ \Delta t \simeq 6.0  \times  10^9 {\rm  yr}/h_{70}$.
Thus we find:

\begin{eqnarray}
{\dot G\over G}\la 36 \times 10^{-12}~h_{70}/{\rm yr}~~&;&~~
\Omega_\Lambda\simeq 0.0  ~,~ \Omega_M\simeq 1.0
 \\
{\dot G\over G}\la 12 \times 10^{-12}~h_{70}/{\rm yr}~~&;&~~
\Omega_\Lambda\simeq 0.8  ~,~ \Omega_M\simeq 0.2
\label{bounds}
\end{eqnarray}

It should be stressed  that these are bounds on ${\dot  G/G}$  around $z
\simeq  0.5$.  Several  local  bounds  on  the  rate  of  change  of the
gravitational  constant,  ${\dot  G_0/G_0}$,  have  been  obtained,  for
example, from binary pulsars, from the Viking Radar and from Lunar Laser
Ranging,  yielding typical upper bounds of ${\dot G_0/G_0} \la 10 \times
10^{-12} {\rm yr}^{-1}$ (see, for instance,  \cite{wil2}).  Other bounds
come from the theory of stellar  evolution,  like  white  dwarf  cooling
\cite{gbea},  being the bounds of the order of $10 \times  10^{-12} {\rm
yr}^{-1}$.  Finally, it should be mentioned that some other local bounds
are as low as  ${\dot  G_0/G_0}  < 6 \times  10^{-12}  {\rm  yr}^{-1}  $
\cite{dw96}.  Note,  however,  that the  values of all these  bounds are
comparable to that obtained  here.  Thus, despite the fact that the SNIa
bounds on $G/G_0$ are quite loose, the longer time baseline  obtained by
using high redshift  measurements  puts stronger  constraints  on ${\dot
G/G}$.  These bounds on the change in $G$ correspond to redshifts  which
have not been  explored yet in the more  standard  gravitational  tests,
mostly  based on solar  system and stellar  physics.  We can combine the
SNIa and local bounds further to set a bound on $\ddot G$:

\begin{equation}
{\ddot G \over G} \la 4 \times 10^{-21}~(h_{70}/{\rm yr})^2
\label{ddotG} 
\end{equation}

In terms of $\omega(z)$  we can also set some further  limits by doing a
Taylor expansion as in \cite{beps99}:

\begin{equation}
\omega^{-1} \simeq \omega^{-1}_0 + z 
\left|{d\omega^{-1}\over{dz}}\right|_0
\end{equation}

\noindent so that we find:

\begin{eqnarray}
\left|{d\omega^{-1}\over{dz}}\right|_0 
\la 1.7
 ~~ &;& ~~ \Omega_\Lambda
\simeq 0.0 ~,~ \Omega_M \simeq 1.0 \\
\left|{d\omega^{-1}\over{dz}}\right|_0 
\la 0.5
~~ &;& ~~ \Omega_\Lambda 
\simeq 0.8  ~,~ \Omega_M \simeq 0.2
\label{boundsdw}
\end{eqnarray}

Future  experiments, such as the Supernovae  Acceleration  Probe (SNAP),
could  achieve a few  percent  magnitude  errors up to  redshifts  of $z
\simeq 1.7$ (see  \cite{albrect}).  By then, we can also  fairly  assume
that other observational data, such as the LSS and CMB experiments, will
provide a  knowledge  of a few  percent on the  cosmological  parameters
\cite{tegmark}.  Thus we can translate the  uncertainty in the magnitude
directly into bounds on $G$:

\begin{equation}
\Delta m \simeq \frac{15}{4}\log\Big(\frac{G}{G_0}\Big).
\end{equation}

Note  that the  possible  effects  of a  varying  $G$ on the  luminosity
distance discussed in the Appendix, are still several times smaller than
the  above  contribution  for  $z\simeq  1.7 $.  For  example,  a  $1\%$
uncertainty in $\Delta m$ (both from peak  luminosity  and  cosmological
parameter  errors) will give us a $0.6\%$  bound in $G/G_0 \la 1.006$ at
$z\simeq  1.7$ or,  equivalently,  $\omega \ga 167$.  The later value is
not  particularly  impressive  as  compared  to the local  bounds in the
context of  Jordan-Brans-Dicke  (JBD)  theories  ($\omega > 2500$).  But
note  that  $z  \simeq  1.7$   corresponds   to  a  look-back   time  in
Eq.[\ref{lookback}]  as large as $ \Delta t \simeq 10  \times  10^9 {\rm
yr}/h_{70}$.  Thus, future data would  eventually  yield firm bounds for
$\dot G$ as low as:

\begin{equation}
{\dot G \over G} \la 6 \times 10^{-13}~h_{70}/{\rm yr}~~;~~~z \simeq 1.7
\label{eq:newGdot}
\end{equation}

\noindent which are more competitive than the current local values.  The
bounds to the change in $\omega$, could be reduced to:

\begin{equation}
\left|{d\omega^{-1}\over{dz}}\right|_0 
\la 3.5 \times 10^{-3} ~~;~~ z \simeq 1.7
\label{eq:newwdot}
\end{equation}

Finally it is  interesting  to mention here that our  analysis  has been
restricted to the peak magnitudes of supernovae.  Comparable  bounds can
be found from the duration of supernovae events  (Eq.[\ref{width}]),  if
the statistical and systematic errors are reduced  significantly,  being
the  advantage  of these  last ones  that are {\sl  independent}  of the
adopted cosmological model.

\section{Discussion and Conclusions}

In astrophysics and cosmology the laws of physics (and in particular the
simplest  version of general  relativity) are  extrapolated  outside its
observational range of validity.  It is therefore  important to test for
deviations  of these laws at  increasing  cosmological  scales and times
(redshifts).  SNIa  provide  us with a new tool to test how the  laws of
gravity and cosmology  were in faraway  galaxies ($z \simeq  0.5$).  The
observational  limits on $\dot G/G$ come from quite different  times and
scales   \cite{wil,wil2,bap},   but  mostly  in  the  local  and  nearby
environments at $z \simeq 0$ (solar system, binary  pulsars, and neutron
stars  \cite{wil2}).  There are also limits derived from the white dwarf
cooling theory  \cite{gbea}, which are based on similar arguments to the
ones  presented in this paper.  Typical  upper bounds give $\dot G/G \la
10^{-11}-10^{-12}$ yr$^{-1}$ ~\cite{wil2}.

Here we have proved by using detailed numerical models that {\sl if} the
value  of $G$ at  the  space-time  location  of  distant  supernovae  is
different  from the local one, it would  change  both the  thermonuclear
energy release and the time scale of the supernova outburst.  The change
can be  quantified  by means of the  change  in the  Chandrasekhar  mass
$M_{\rm  Ch}\propto  G^{-3/2}$, and our detailed  numerical results have
been  interpreted  in terms of a very  simple  physical  model.  To this
regard it is important  to realize  that our  conclusions  would  remain
unchanged should a modification of the parameters of the explosion 
lead to a smaller mass of $^{56}$Ni  synthesized in the supernova event,
leading to a dimmer supernova.

We have also shown in a  self-consistent  way that for plausible  models
for a varying $G$, such as  scalar-tensor  theories, the possible effect
of a varying $G$ on the cosmological  evolution yields a contribution to
the luminosity distance relation which is one order of magnitude smaller
than  the  effect   produced  by  the  same  variation  of  $G$  on  the
Chandrasekhar  mass.  Thus our  approach  complements  the  analysis  of
\cite{ep01,beps99,ss01}  which  neglected  the effects in the physics of
the supernovae and used the evidence for the accelerating  universe as a
way to constraint cosmic evolution in non-standard  theories of gravity.

In this paper we have also found further bounds for a varying $G$ from a
likelihood analysis of the peak luminosities of the Supernova  Cosmology
Project.  Our   results   are   summarized   in  Figures  3  and  4  and
Eq.[\ref{boundl0}]-[\ref{boundk0}],  with values of $G/G_0 \la 1.2$.  We
have  further  translated  these  results  into bounds for $\dot G/G$ in
Eq.[\ref{bounds}],    $\ddot    G   /G$    in    Eq.[\ref{ddotG}]    and
${d\omega^{-1}/{dz}}$  in Eq.[\ref{boundsdw}].  Some of these bounds are
new or comparable to other existing  estimates from the local  universe,
which typically gives stronger  constraints  for $\omega$ or $G/G_0$, at
least within JBD models.

In the  context of JBD or STT  models the limits we find for $\dot  G/G$
correspond to $\omega \ga 3-30$ and are therefore less restrictive  than
the solar system limits  $\omega \ga 2500$  \cite{wil2}.  However,  STTs
could allow for  $\omega=\omega(\phi)$.  To be precise,  $\omega$ is not
required to be a constant, so that $\omega$  could  increase with cosmic
time,  $\omega=\omega(z)$,  in such a way  that it  could  approach  the
general relativity  predictions ($\omega \rightarrow \infty$) at present
time and still give significant deviations at earlier cosmological times
\cite{ep01,beps99,ss01}.  Furthermore, it has been shown \cite{dan} that
the cosmological evolution makes STTs practically indistinguishable from
General  Relativity  at  the  present  epoch.  Our  results  set  strong
constraints at cosmological distances.

The  interest  of these new  bounds  with  respect  to the other  values
discussed  so far in the  literature,  is not  whether  or not  they are
better, but the facts that:  i) a different  method has been  tested and
used and, ii) our bounds correspond to higher redshifts, $z \simeq 0.5$,
thus  extending the  constrains on the evolution of $G$ all the way from
solar/stellar  distances  to Gpc,  that is by more  than  15  orders  of
magnitude.  In this  sense,  cosmological  nucleosynthesis  also  offers
another  limit on the amount of variation  of $G$.  Generally  speaking,
the  bounds  derived  from  primordial  nucleosynthesis  arise  from the
sensitivity  of the  abundances  of  light  elements  produced  at  high
temperature to the expansion rate of the Universe at those temperatures,
especially  $^4$He.  There is a range of opinions, but there is also the
widespread  agreement that the expansion rate must have been well within
a factor of two of the  standard  model.  Some might  even push for more
stringent  limits that would exclude  changes by even as little as $\sim
10$\%, which would be  marginally  consistent  with our analysis --- see
the  most  recent  analysis  presented  in  \cite{sea}  for  a  detailed
discussion.

Finally,  we would  like to  stress  that new  observations  of  distant
supernovae, or other standard candles, at higher redshifts  ($z>1$) will
constrain  even  more  the  current  limits  on  the  variation  of  the
fundamental constants (see  Eq.[\ref{eq:newGdot}-\ref{eq:newwdot}]).  To
this regard it is important to realize that the recently  analyzed  SNIa
1997ff  \cite{gnp}, the oldest and most distant SNIa ever  discovered at
$z \simeq  1.7$  \cite{r17},  could  provide  an  important  test of the
viability of alternative theories of gravity.

\begin{acknowledgements}
This work has been supported by the DGES grant  PB98--1183--C03--02,  by
the   MCYT   grants    AYA2000--1785,    AYA2000--1574,    BFM2000-0810,
ESP199-1803-E  and ESP98--1348 and by the CIRIT grants  1995SGR-0602 and
2000ACES-00017.  One of us, EGB, also  acknowledges the support received
from   Sun   MicroSystems    under   the   Academic    Equipment   Grant
AEG-7824-990325-SP.
\end{acknowledgements}

\appendix

\section{Scalar-Tensor Theories}

The main  topic of this  paper  is how the  Hubble  diagram  of  distant
supernovae  could help to set constraints on a varying $G$.  In order to
do that we need to study the  physics of SNIa, but to be  consistent  we
need to derive a luminosity  distance  relation  which also includes the
possible  effects of a varying $G$ on the  cosmological  evolution  (see
\cite{ls72}).  In this  Appendix we will show that for a given change in
$G$ this later effect is typically  smaller  than the one induced by the
change in the Chandrasekhar mass in the energy release of SNIa.

The  possibility  that $G$ could  vary in space  and/or  time  naturally
appears in the  framework of  Scalar-Tensor  theories of gravity  (STTs)
such  as JBD  theory  or its  extensions.  These  models  have  recently
attracted a large interest (see \cite{ep01} and references therein).  To
make quantitative predictions we will consider cosmic evolution in STTs,
where $G$ is derived from a scalar field $\phi$  which is  characterized
by a function  $\omega=\omega(\phi)$ that determines the strength of the
coupling  between the scalar  field and  gravity.  In the  simplest  JBD
models, $\omega$ is just a constant and $G \simeq \phi^{-1}$, however if
$\omega$   varies  then  it  can  increase  with  cosmic  time  so  that
$\omega=\omega(z)$.  The Hubble rate $H$ in these models is given by:

\begin{equation}
H^2 \equiv \left({{\dot a}\over{a}}\right)^2= {8 \pi \rho\over{3\phi}}
+{1\over{a^2 R^2}}+ {\Lambda\over{3}}+{\omega\over{6}}{{{\dot \phi}^2}
\over{\phi^2}} - H {{{\dot \phi}}\over{\phi}},
\label{H^2}
\end{equation}

\noindent  this equation has to be  complemented  with the  acceleration
equations  for $a$ and  $\phi$,  and with the  equation  of state  for a
perfect fluid:  $p=(\gamma-1)\rho$  and ${\dot  \rho}+3\gamma  H\rho=0$.
The  structure of the  solutions to this set of equations  is quite rich
and  depends   crucially  on  the   coupling   function   $\omega(\phi)$
\cite{bap}).  Here  we are  only  interested  in  the  matter  dominated
regime:  $\gamma=1$.  In the weak field  limit and a flat  universe  the
exact solution is given by:

\begin{equation}
G= {4+2\omega\over{3+2\omega}}\phi^{-1} = G_0 (1+z)^{1/(1+\omega)}.
\label{G(t)}
\end{equation}

\noindent  In this  case we also  have  that  $a =  (t/t_0)^{(2\omega+2)
/(3\omega+4)}$.  This  solution for the flat  universe is recovered in a
general case in the limit $t  \rightarrow  \infty$ and also arises as an
exact solution of Newtonian  gravity with a power law $G \propto  t^{n}$
\cite{bar}.  For non-flat  models, $a(t)$ is not a simple  power-law and
the solutions get far more  complicated.  To illustrate the effects of a
non-flat  cosmology  we will  consider  general  solutions  that  can be
parameterized as Eq.[\ref{G(t)}]  but which are not simple power-laws in
$a(t)$.  In this case, it is easy to check that the new Hubble law given
by Eq.[\ref{H^2}] becomes:

\begin{equation}
H^2 = H^2_0 ~\left[ ~\hat\Omega_M (1+z)^{3+1/(1+\omega)} +
\hat\Omega_R (1+z)^2 + \hat\Omega_\Lambda ~\right]
\label{H(z)}
\end{equation}

\noindent where  $\hat\Omega_M$,$\hat\Omega_R$  and $\hat\Omega_\Lambda$
follow  the  usual   relation:  $\hat   \Omega_M+  \hat  \Omega_R+  \hat
\Omega_\Lambda  = 1$ (an overall factor would just redefine the value of
$H_0$) and are related to the familiar local ratios ($z \rightarrow 0$):
$\Omega_M \equiv 8\pi G_0 \rho_0 /(3H_0^2)$, $\Omega_R =1/(R H_0)^2$ and
$\Omega_\Lambda =\Lambda/(3H_0^2)$ by:

\begin{eqnarray}
\hat\Omega_M&=&\frac{\Omega_M}{g}
~\left(\frac{3+2\omega}{4+2\omega}\right)   ~ ; ~
\hat\Omega_\Lambda=\frac{\Omega_\Lambda}{g} ~ ; ~
\hat\Omega_R=\frac{\Omega_R}{g}\\
g &\equiv& 1+\frac{1}{(1+\omega)}-\frac{1}{6}~\frac{\omega}
{(1+\omega)^2}.
\label{omegahat}
\end{eqnarray}

Thus the general  relativity  limit is recovered as $\omega  \rightarrow
\infty$.  For a  flat  universe,  the  luminosity  distance  $d_L=d_L(z,
\Omega_M,   \Omega_\Lambda,  \omega)$  is  related  the  (line-of-sight)
comoving coordinate distance $r(z)$ as:

\begin{equation}
d_L= {r(z)\over{a}} = {c\over{H_0}} (1+z) ~\int   {dz' \over{H(z')}}.
\end{equation}

In  the  general  case  we  have  to  replace  the  integral   with  its
trigonometric   or  the  hyperbolic   sinus  to  account  for  curvature
\cite{pbs}.  In the  limit of small  $z$ we  recover  the  usual  Hubble
relation:  $d~H_0/c = z-(1+\hat  q_0)  z^2/2$  where a new  deceleration
$\hat q_0$ parameter is related to the standard one by:

\begin{equation}
\hat q_0 ~=~{q_0\over{g}} + {\hat\Omega_M\over{2(1+\omega)}}.
\end{equation}

\noindent One can see from these  equations that even for relative small
values of $\omega$  the effect of a varying  $G$ on $d_L$ is small.  For
example for the flat case  ($\hat\Omega_R=0$  and  $\hat\Omega_M=1$)  at
$z=0.5$  we have  $d_L=0.5456$  for  $\omega=10$  and  $d_L=0.5505$  for
$\omega=10^3$ ($d_L$ in units of $c/H_0$).  Thus, the change in $\omega$
produces  brighter  apparent  objects, in this case  $\Delta m= -0.019$,
which would tend to partially  compensate the dimmering produce by the a
varying $G$ on the Chandrasekhar mass:  in this case $\Delta m= +0.060$.
In general, we find that the  cosmological  effect in the Hubble diagram
of SNIa is always smaller, by factors of a few, than the effect produced
by a varying $G$ on the  Chandrasekhar  mass.  Also in the general case,
the  cosmological  evolution in a model with  increasing $G$ at high $z$
tends to  decrease  the  acceleration  (with  respect  to the case  with
constant $G$), which partially  compensates the apparent increase due to
the the change in the Chandrasekhar mass.

\end{document}